\documentclass[prl,twocolumn]{revtex4}
\usepackage{graphicx}
\usepackage{amssymb}
\usepackage{amsmath}

\graphicspath{{figures/}}

\begin{document}
\title{Narrow-line magneto-optical cooling and trapping of strongly magnetic atoms}
\author{Andrew J. Berglund${}^1$}
\author{James L. Hanssen${}^{1,2}$}
\author{Jabez J. McClelland${}^1$}\affiliation{$^{1}$Center for Nanoscale Science and Technology,
National Institute of Standards and Technology, Gaithersburg, MD
20899\\${}^2$Maryland NanoCenter, University of Maryland, College
Park, MD 20742}
\date{\today}
\begin{abstract}
Laser cooling on weak transitions is a useful technique for reaching
ultracold temperatures in atoms with multiple valence electrons.
However, for strongly magnetic atoms a conventional narrow-line
magneto-optical trap (MOT) is destabilized by competition between
optical and magnetic forces. We overcome this difficulty in Er by
developing an unusual narrow-line MOT that balances optical and
magnetic forces using laser light tuned to the blue side of a narrow
(8~kHz) transition. The trap population is spin-polarized with
temperatures reaching below $2~\mu$K. Our results constitute an
alternative method for laser cooling on weak transitions, applicable
to rare-earth-metal and metastable alkaline earth elements.
\end{abstract}
\maketitle

The extension of laser cooling techniques to weak, narrow-line
optical transitions \cite{Castin:1989a, Wallis:1989a} opens new
possibilities to reach low temperatures and high phase-space
densities for atoms with multiple valence electrons
\cite{Katori:1999a}. The main advantage of a kilohertz-linewidth
($\Gamma/2\pi\approx1~$kHz) transition is the low Doppler
temperature, $\hbar \Gamma/(2 k_B)$, which in principle enables
Doppler laser cooling to nanokelvin temperatures without evaporative
techniques. However, the maximum radiation pressure force that can
be applied, $\hbar k\Gamma/2$, is correspondingly small for a narrow
transition and may not even be sufficient to levitate atoms against
gravity ($2\pi/k$ is the optical wavelength). Experimental studies
of narrow-line laser cooling have concentrated on the ${}^1$S${}_0$
to ${}^3$P$_1$ transitions in Mg, Ca, and Sr. For Sr
\cite{Katori:1999a, Mukaiyama:2003a:PhysRevLett.90.113002,
loftus:2004a:073003, loftus:2004b, nagel:2005a:083004}, optical
forces are sufficient to compensate gravity, but the resulting
interplay of forces modifies the trap dynamics
\cite{loftus:2004a:073003,loftus:2004b}. For Ca and Mg, optical
forces are weaker ($\Gamma$ is smaller), so that more complicated
schemes such as quenched cooling
\cite{Binnewies:2001a:PhysRevLett.87.123002, Curtis:2001a},
metastable trapping \cite{Grunert:2002a:PhysRevA.65.041401}, and
two-photon cooling \cite{malossi:2005a:051403} are necessary for
stable trapping against gravity.

The rare-earth-metal element Er also has multiple valence electrons
and narrow optical transitions \cite{Ban:2005a}, but in contrast to
the alkaline earths, Er also has a large ground state magnetic
moment ($7~\mu_B$) that provides an additional ``handle" for
manipulation. For the narrow ($\Gamma/2\pi=8~$kHz) transition at
841~nm in Er, the magnetic force is equal to the maximum radiation
pressure force at a field gradient of only 0.3~T/m. As a result, a
magneto-optical trap (MOT) is destabilized if the trapping lasers
tend to drive atoms into untrapped magnetic sublevels. In this
Letter, we exploit this comparable magnitude of optical and magnetic
forces in order to form a narrow-line MOT for Er. In our ``strongly
magnetic" MOT, confinement is provided by a combination of
dissipative optical forces and conservative magnetic forces. The
trap is globally stable when the trapping lasers are tuned to the
\emph{blue} side of the (unshifted) atomic resonance. The resulting
trap forms at a position of non-zero magnetic field, with a
spin-polarized population at temperatures below 2~$\mu$K. In a
conventional MOT, on the other hand, confinement forces are purely
optical, cooling occurs for red-detuned lasers, and the trap
population is (on average) unpolarized at a position of zero
magnetic field. The unusual features of our trap are explained by a
Doppler cooling model including optical pumping between magnetic
sublevels. Our results constitute the lowest temperature achieved
for Er atoms and provide an important benchmark in the study of
highly magnetic rare-earth-metal elements.

Magnetic forces modify the description of magneto-optical cooling
and trapping in a nontrivial way. Consider the $z$ motion of an atom
in a magnetic field $B(z)=B' z$ ($B'>0$), together with a circularly
polarized traveling-wave laser beam (frequency $\omega_L$, wave
vector $\vec{k}=k \hat{z}$ with positive or negative $k$
corresponding to the direction of propagation) detuned by $\delta =
\omega_L-\omega_A$ from the atomic transition $\omega_A$. We
restrict attention to $z>0$ where $B(z)$ is positive. Including the
gravitational force $-Mg\hat{z}$, the semiclassical position $z$-
and velocity $v_z$-dependent force on the atom is \cite{Lett:1989a,
Metcalf:1999,loftus:2004a:073003,loftus:2004b}
\begin{equation}\label{Eq:Force1D} F_z(z,v_z) =F_0 +\frac{\hbar k \Gamma}{2}
\frac{s}{1+s +4\left(\delta/\Gamma - z/z_0^\pm -
kv_z/\Gamma\right)^2}\end{equation} where
$$F_0 = -g_g m_J\mu_B  \left|B'\right|-M
g\quad,\quad z_0^\pm = \frac{\hbar\Gamma}{\mu'_\pm |B'|}.$$
$\mu'_\pm= (\pm g_e+m_J\Delta g_{eg})\mu_B$, $g_g$ and $g_e$ are the
Land\'{e} factors for the ground and excited states, $\Delta
g_{eg}=g_e-g_g$, $m_J$ is the projection of the angular momentum $J$
along the local magnetic field direction, $s$ is the saturation
parameter \cite{Metcalf:1999}, and $\pm$ corresponds to $\sigma^\pm$
polarization.

The force field of Eq.~\eqref{Eq:Force1D} is characterized by the
parameter $\mathfrak{F}_0 = 2 F_0/(\hbar k\Gamma)$, the ratio of the
magnetic and gravitational forces to the light force. For $k<0$,
$\sigma^-$ polarization, and red laser detuning $\delta<0$,
simultaneous cooling and trapping (\emph{i.e.}, stable equilibrium
in both $z$ and $v_z$) are achieved for
$0<\mathfrak{F}_0<s/(1+s)\leq 1$, which also requires $m_J<0$. Near
equilibrium, the force is approximately $F_z(z,v_z) \approx -\kappa
(z-z_{eq})-\beta v_z$ where $z_{eq}$ is the trap position, and
$\kappa$ and $\beta$ define the spring constant and velocity damping
coefficient, respectively. At $z_{eq}$, magnetic forces push the
atom away from the origin while optical forces push it back toward
the origin, and the red-detuned laser provides Doppler cooling
towards $v_z=0$. While cooling and trapping may be expected for a
red-detuned beam, a similar equilibrium point exists for $\delta>0$,
$\sigma^+$ polarization, and $k>0$ in the weak-field seeking states
with $m_J>0$. In this case, magnetic forces pull the atom toward the
origin while optical forces push it away and the laser still
provides Doppler cooling because it is red-detuned from the
Zeeman-shifted transition frequency at the trap position. Therefore,
a stable trap forms for appropriately polarized red- \emph{or}
blue-detuned laser light, as long as the direction of laser
propagation opposes $F_0$ and the radiation pressure is sufficiently
large.

\begin{figure}[t]\includegraphics[width=3.3in]{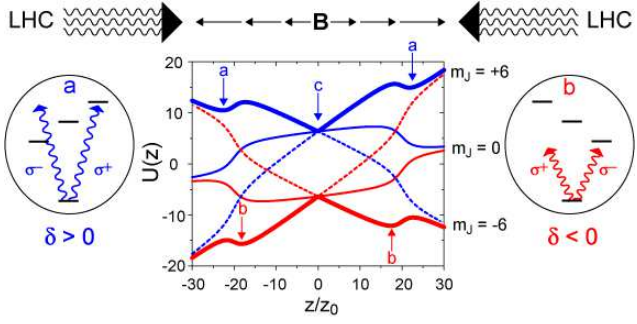}
\caption{\label{Fig:NarrowLineTheory}(Color online) Potential energy
for a magnetic atom in a narrow-line MOT. $U(z)$ is in units of
$\hbar k\Gamma z_0/2$. At the top, counterpropagating laser beams
are labeled by their left-hand circular (LHC) helicity. Curves are
shown for an atom with $J=6$, for red or blue detuning and various
values of $m_J$. For an atom in state $m_J$, the curve at higher
(lower) energy [blue (red) online] represents the potential for
$\delta>0$ ($\delta <0$). Optical pumping stabilizes the stretched
trap states ($m_J=\pm J$), since blue (red)-detuned trapping lasers
tend to drive atoms toward the upper (lower) potential energy curve.
Increasing line thickness (from dashed to thin to thick lines)
represents increasing population of a state due to optical pumping.
Insets a) and b) show the resonant optical transitions at trap
minima labeled $a$ and $b$. Here, for $m_J=J$, $\mathfrak{F}_0=1/2$
in the absence of gravity; the gravitational force is 20~\% of the
magnetic force.}\end{figure}

Now consider counterpropagating laser beams carrying opposite
angular momenta, as in a standard MOT \cite{Metcalf:1999}. In this
situation, the sign of the magnetic force reverses at $z=0$, and the
same circular polarization induces $\sigma$ transitions of opposite
sign. As a result, stable equilibria form on either side of the
origin, and a conservative magnetic trap also forms at $z=0$ for
weak-field seeking ($m_J>0$) substates. Generalizing
Eq.~\eqref{Eq:Force1D} to include the changing sign of the magnetic
field and summing the force of each beam, we find the potential
energy curves $U(z) = -\int F_z(z',0)dz'$ shown in
Fig.~\ref{Fig:NarrowLineTheory}. For the non-magnetic $m_J=0$
substate with $\delta<0$, the potential energy $U(z)$ reproduces the
results in Refs.~\cite{loftus:2004a:073003, loftus:2004b} for a
far-detuned, narrow-line Sr MOT in the presence of gravity. For
magnetic states ($m_J\neq 0$) in the absence of gravity, the
potential minima at points labeled $a$ and $b$ exhibit identical
values of $\kappa$ and $\beta$. Point $c$ is a purely magnetic trap.
Optical pumping populates the ``stretched" trap states $|m_J|=J$ in
a detuning-dependent way. Blue-detuned trapping lasers become
resonant on $\sigma^+$ transitions and therefore tend to drive atoms
to the $m_J=+J$ magnetic substate exhibiting a globally stable,
V-shaped trap potential. In contrast, red-detuned lasers approach
resonance on $\sigma^-$ transitions tending to drive atoms to the
$m_J=-J$ substate, with a globally unstable $\Lambda$-shaped
potential. We thus expect, and find in practice, that the
blue-detuned trap is quite robust, while we were unable to observe
any trap in the red-detuned case. Note also that magnetic forces can
compensate gravity in this trap, and thus \emph{optical forces need
not exceed gravitational forces}. This trap may therefore be useful
for narrow-line cooling of metastable, magnetically trapped alkaline
earth atoms \cite{Grunert:2002a:PhysRevA.65.041401} in cases where
$\hbar k\Gamma/2 \lesssim Mg$, or even for cooling on ultranarrow
transitions such as the 2.1~Hz transition in Er \cite{Ban:2005a}.

\begin{figure}[t]\includegraphics*[width=3.2in]{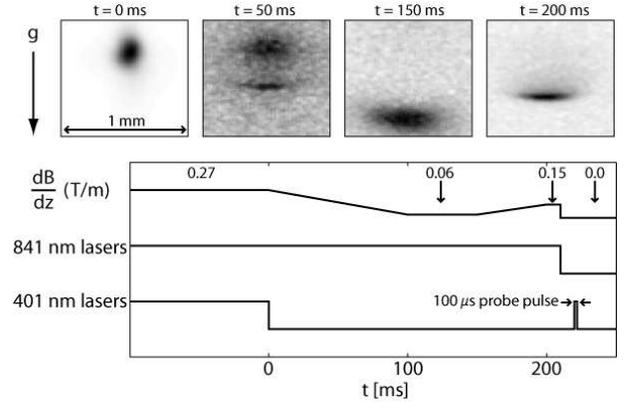}
\caption{\label{Fig:LoadingProcedure}Cloud images (top) and timing
sequence (bottom) during trap loading.}\end{figure}

To investigate the cooling limits of this force, we note that near
$z_{eq}$ the kinetic energy of an atom increases by twice the photon
recoil energy ($\hbar^2k^2/2M$) at the equilibrium absorption rate
$\mathfrak{F}_0\Gamma/2$, while the energy damping rate is $\beta
v_z^2$ \cite{Lett:1989a}. These heating and damping rates are equal
for a kinetic energy distribution corresponding to a temperature
\begin{equation}\label{Eq:Teq} T_{eq} = T_D \frac{s/2}{
\mathfrak{F}_0\sqrt{s/\mathfrak{F}_0-1-s}}\quad,\quad
T_D=\frac{\hbar \Gamma}{2 k_B}.\end{equation} Recalling the
stability constraint that $0<\mathfrak{F}_0<s/(1+s)$, we find that
$T_{eq}$ reaches a minimum of $T_D$ for a very dilute,
spatially-extended trap, \emph{i.e.}, for $\mathfrak{F}_0\ll 1$.
However, one is usually interested in creating a compact sample that
is not only \emph{cold} but also \emph{dense}. Neglecting gravity
for simplicity, we find that the 1D phase-space
density is maximized for $\mathfrak{F}_0 = 1/2$. For the 841~nm
transition in Er, the highest phase-space density is predicted
to occur at a reasonable magnetic field gradient of 0.15~T/m with
1D kinetic energy distribution corresponding to
$T_{eq}=2T_D=380$~nK.

To observe these dynamics, we modified the apparatus in
Refs.~\cite{mcclelland:2006a:143005, Berglund:2007b}. In brief,
approximately $10^5$ $^{166}$Er atoms are trapped and cooled to
150~$\mu$K in a standard MOT on the bright ($\Gamma/2\pi=36$~MHz
\cite{mcclelland:2006b:064502}) transition at 401~nm. For
narrow-line cooling, we use an 841~nm diode laser locked by a fast
servo to a passively stable cavity \cite{Fox:2002a}. Acousto-optic
modulators are used for fine control of both the frequency and
power. At the MOT position, we use a maximum 841~nm laser power of
3~mW distributed into three orthogonal beams with Gaussian waists of
($4.5\pm0.05$)~mm \footnote{Quoted errors and error bars represent
one standard deviation combined random and systematic uncertainty.}.
The 401~nm beams are tilted by 6$^\circ$ to provide optical access
for the 841~nm beams. Temperature measurements are made by turning
off all optical and magnetic fields then imaging fluorescence from
the expanding cloud by pulsing on the 401~nm MOT lasers for
100~$\mu$s.

\begin{figure}[t]\includegraphics[width=3.in]{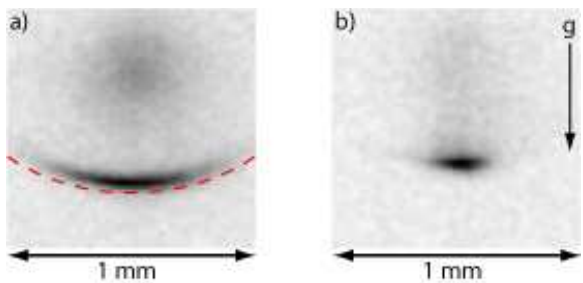}
\caption{\label{Fig:NarrowLineMOT}(Color online) Narrow-line MOT
images. a) Cooling and trapping with a \emph{single} downward
propagating beam. Atoms collect along an ellipse of constant
magnetic field (dashed line) where they come into resonance with the
laser. b) Full 3D narrow-line MOT. In both images, a faint cloud of
atoms trapped in the conservative magnetic trap can be seen as
well.}\end{figure}

The narrow-line MOT loading procedure is shown in
Fig.~\ref{Fig:LoadingProcedure}. Atoms are initially trapped and
cooled on the 401~nm transition (the 841~nm beams are also on during
this period). At time $t=0$, the 401~nm beams are extinguished leaving only the 841~nm beams. The
field gradient is ramped down over 100~ms to allow atoms to collect
in the narrow-line MOT, then raised to recompress the atomic cloud.
Two typical narrow-line MOT fluorescence images are shown in
Fig.~\ref{Fig:NarrowLineMOT}. In image a), we used a \emph{single}
blue-detuned ($\delta/2\pi\approx 1$~MHz) beam to push atoms into a
1D narrow-line MOT. Because the atoms are polarized in magnetically
trapped $m_J>0$ substates, they are confined by magnetic forces
along $\pm x$, $\pm y$ and $+z$ and an optical force along $-z$.
Cloud expansion measurements show asymmetric velocities
corresponding to temperatures $T_x=(48\pm3.3)~\mu$K along the $x$
direction and $T_z=(3.9\pm0.4)~\mu$K along the $z$ direction,
indicating cooling along the direction of laser propagation. In
image b), we used all three retroreflected beams. Here, we see
cooling and confinement in all directions with more symmetric expansion
velocities corresponding to $T_x=(4.8\pm2.1)~\mu$K and
$T_z=(1.2\pm0.3)~\mu$K.

\begin{figure}[t]\includegraphics[width=3.3in]{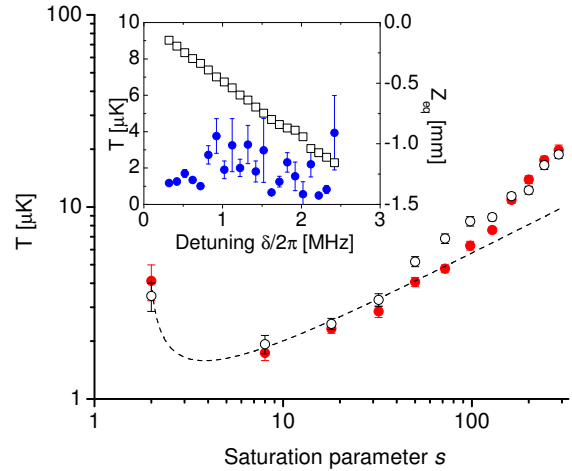}
\caption{\label{Fig:TvsS}(Color online) Narrow-line MOT temperatures
$T_x$ (open) and $T_z$ (filled) as a function of the axial-beam
saturation parameter $s$. The dashed curve is a fit to
Eq.~\eqref{Eq:Teq} for $s \leq 100$, with fit parameters
$T_D=540$~nK and $\mathfrak{F}_0=0.65$, to be compared with the
expectations $T_D=180$~nK and $\mathfrak{F}_0=0.3$. Inset:
Temperature $T_z$ (circles, left axis) and trap position $z_{eq}$
(squares, right axis) vs. detuning $\delta/2\pi$.}\end{figure}

We measured the temperature of the 3D narrow-line MOT as a function
of $s$ and $\delta$, for a ratio of (total) transverse to axial beam
power of 0.44. A number of predictions of the 1D model are confirmed
in Fig.~\ref{Fig:TvsS}. In particular, the temperature exhibits a
minimum for a small value of $s$ and increases as $\sqrt{s}$ for
$s\leq 100$, the temperature is largely independent of $\delta$, and
the trap position varies linearly with $\delta$. (Note the similar
behavior in Refs.~\cite{loftus:2004a:073003,loftus:2004b} where
red-detuned lasers balanced Sr atoms against gravity).

\begin{figure}[t]\includegraphics[width=3.3in]{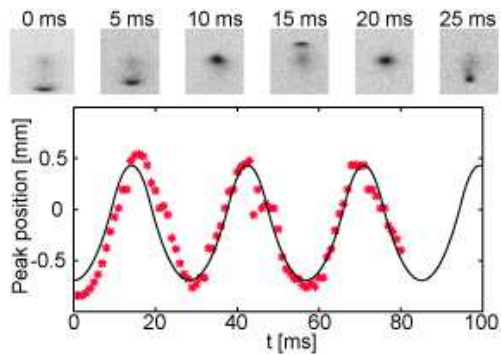}
\caption{\label{Fig:Revival}(Color online) Oscillation of the atomic
cloud after release into the magnetic trap. (Upper) Cloud images at
early times. (Lower) $z$ position of maximum density together with a
fit to the expected trajectory.}\end{figure}

To test the spin polarization, we turn off the 841~nm lasers at
$t=0$, leaving all magnetic fields on. Because atoms are in
weak-field seeking states, they feel a restoring force toward the
origin. As shown in Fig.~\ref{Fig:Revival}, atoms begin oscillating
in the conservative potential with a period of approximately 30~ms.
We observe high-fidelity oscillations in the position of peak
density, with the oscillation period indicating a spin polarization
of $(74\pm6)~\%$. After a few oscillation periods, the anharmonic
quadrupole potential leads to complicated cloud trajectories
\footnote{See EPAPS for a movie of the cloud oscillation for 100~ms
after release into the magnetic trap. For more information on EPAPS,
see http://www.aip.org/pubservs/epaps.html.}.

Our trap is relatively insensitive to the polarization of the
transverse beams, and forms even when some retroreflections are
blocked. In contrast, the trap is very sensitive to the polarization
of the vertical trapping laser, while the retroreflection of this
beam is inconsequential. We have formed traps above and below the
origin by reversing the direction of a single beam. The lower trap
is more robust, presumably due to the stabilizing influence of
gravity. These trap characteristics can be qualitatively explained
from geometric considerations. Because of the large detuning and
narrow atomic line, atoms approach resonance only within a narrow
ellipsoid of constant magnetic field
\cite{loftus:2004a:073003,loftus:2004b}. $\pi$ and $\sigma^-$
transitions can be neglected due to the large value of $\delta$, so
the polarizations of the transverse beams only affect the coupling
efficiency to the $\sigma^+$ transition, via a position-dependent
geometrical factor. Although the overall behavior of the trap is
well-understood, our measured temperatures are always slightly
higher than predicted by the simple 1D model. This discrepancy may
be explained by heating during temperature measurement, since we
observe that transients during magnetic field shutoff impart
residual cloud velocities of a few cm/s. Given the total energy
required to accelerate the cloud, a few percent spread in spin
polarization results in a few percent spread in cloud velocity,
corresponding to $\mu$K-scale heating. The cloud temperature may
also be affected by stimulated transitions or polarization-gradient
effects, which cannot be properly analyzed without a 3D quantum
mechanical model.

In summary, we have demonstrated a new type of narrow-line MOT for
strongly magnetic atoms, in which optical and magnetic forces play
equally important roles. In future experiments, we plan to optimize
loading procedures in order to study the atom number and phase-space
density for this trap. These techniques should be applicable to the
other rare-earth-metal elements and to alkaline earth elements in
magnetically trapped, metastable states.

The authors gratefully acknowledge contributions from S.~A. Lee.
This work has been supported in part by the NIST-CNST/UMD-NanoCenter
Cooperative Agreement. A.~B. acknowledges financial support from the
National Research Council.


\end{document}